\documentclass[a4paper,aps,onecolumn,showpacs,showkeys,nofootinbib,preprintnumbers,superscriptaddress,amsmath,amssymb,amsfonts]{revtex4}
\usepackage{graphicx}
\usepackage{bm,natbib}
\usepackage{amsmath}
\usepackage[english]{babel}
\usepackage[T1]{fontenc}
\usepackage{amssymb}
\usepackage{amsfonts}
\usepackage{epsfig}
\usepackage{colordvi}
\usepackage{psfrag}
\usepackage{color}
\usepackage{ mathrsfs }

\newcommand{\Lagr}{\mathcal{L}}

\newcommand{\G}{\mathcal{G}}

\usepackage{dcolumn}
\usepackage{multirow}
\usepackage{hyperref}
\usepackage{epstopdf}
\usepackage{bm}
\usepackage{times}

\hypersetup{
  colorlinks=true,        
  linkcolor=blue,         
  citecolor=magenta,      
}

\begin{document}
\title{Equivalence  of non-minimally coupled cosmologies  by Noether symmetries}


\author{Francesco Bajardi}, 
\affiliation{Universit\'a di Napoli "Federico II", Compl.~Univ.~di Monte S.~Angelo, Edificio G, Via Cinthia, I-80126, Napoli, Italy.}
\affiliation{INFN Sezione  di Napoli, Compl. Univ. di
Monte S. Angelo, Edificio G, Via Cinthia, I-80126, Napoli, Italy.}

\author{Salvatore Capozziello}
\affiliation{Universit\'a di Napoli "Federico II", Compl.~Univ.~di Monte S.~Angelo, Edificio G, Via Cinthia, I-80126, Napoli, Italy.}
\affiliation{INFN Sezione  di Napoli, Compl. Univ. di
Monte S. Angelo, Edificio G, Via Cinthia, I-80126, Napoli, Italy.}
\affiliation{Laboratory of Theoretical Cosmology,Tomsk State University of Control Systems and Radioelectronics(TUSUR), 634050 Tomsk, Russia.}

\date{\today}

\begin{abstract}
We discuss   non-minimally coupled cosmologies involving different geometric invariants. Specifically,  actions containing  a  non-minimally coupled scalar field to gravity described, in turn, by   curvature,  torsion and  Gauss--Bonnet scalars are considered.  We show that couplings, potentials and kinetic terms are determined by the existence of Noether symmetries which, moreover, allows to  reduce and solve dynamics.   The main finding of the paper is that different non-minimally coupled theories, presenting the same Noether symmetries, are dynamically equivalent. In other words, Noether symmetries are a selection criterion to compare different theories of gravity. 
\end{abstract}


\maketitle
\section{Introduction}
The Hilbert-Einstein action, linear in the Ricci curvature scalar $R$,  gives rise to the field equations of  General Relativity (GR) which is the theory of gravity capable of  fitting a huge amount of phenomena ranging from gravitational waves, astrophysical compact  objects, black holes up to  cosmology. Despite of this large bulk of positive results, several shortcomings must be overcome by GR in order to match the whole budget of astrophysical and cosmological  observations and to fix inconsistencies    at  quantum level. In other words, we lack a theory of gravity addressing the phenomenology, in a self-consistent way,  from ultraviolet to infrared scales.

In particular, at large scales,  discrepancies occur between the theoretical value of the Hubble constant predicted by GR and the  measured one  \cite{Addison:2015wyg}; furthermore, without including \emph{dark Matter} or \emph{dark energy}, the theory is unable to explain the today observed accelerated behavior of the  universe \cite{Riess:1998cb, Bamba:2012cp,Copeland}, the missing matter in galaxies and the structure formation \cite{Read:2004xc}. Furthermore, the "local" formulation of GR seems completely in disagreement with the intrinsic "non-locality" of quantum mechanics.

Besides  the today observed cosmological acceleration,  also the early epoch inflation cannot be framed considering only GR. Starting from 1980,  A.A. Starobinsky \cite{Starobinsky},  A. Guth  \cite{Guth:1980zm},  A. Linde  \cite{Linde:1983gd}, et al.,  proposed  possible solutions for this problem:  modifications of the gravity action containing minimal and non-minimal  couplings between geometry and  scalar fields or the introduction of  higher-order curvature terms, behaving like  scalar fields, \emph{e.g.} the so called {\it scalaron}, have been taken into account. See \cite{Bamba:2015uma} for a review. 

The paradigm is that  introducing a new field, a generic \emph{dilaton}, shortcomings of  Standard Cosmological Model can be  solved. These modified actions, subsequently, have been studied also for  late-time cosmology \cite{Sadjadi:2013nb}, for black holes \cite{Gibbons:1987ps}, for string-dilaton cosmology considering  different kinds of couplings and potentials \cite{Capozziello:1993vs, Capozziello:1993ts}.  An important feature of these models is that, under suitable conformal transformations, they can be always  reduced to GR plus scalar fields.
The frame containing  non-minimally couplings  and higher-order terms is the \emph{Jordan Frame}, while the standard GR  plus scalar fields is restored in  the  \emph{Einstein Frame}.

As a general remark, with the purpose of addressing or,  at least, alleviating the above problems, alternative  theories of gravity have been proposed. The  approach is   considering effective Lagrangians where  further geometric invariants and/or scalar fields are included. For example, relaxing the hypothesis of second-order field equations, we may introduce into the action a function of the Ricci scalar like $f(R)$ \cite{Sotiriou:2008rp, Capozziello:2011et}, of its derivatives \cite{Amendola:1993bg, Gottlober:1989ww}, or  include other curvature invariants adopting a metric or metric-affine formulation of the theory \cite{Allemandi}. Furthermore, in the context of   GR extensions, non-minimal couplings between gravity and  dynamical scalar fields are particularly useful for dynamical reasons. 
Among the second order curvature invariants, there are also models considering    the Gauss--Bonnet topological invariant $\G$ adopted  to obtain ghost-free dynamics. For general reviews on modified gravity see, for example,  \cite{Nojiri:2017ncd, Nojiri:2010wj, Nojiri:2017qvx}.

 Moreover, also affine connection can play a key role in modifications and extensions of standard GR. Levi-Civita connection, for instance, represents only the simplest case of torsionless affine connection and, without discarding the anti-symmetric part of $\Gamma^\alpha_{\mu \nu}$,   \emph{torsion} naturally arises. In this framework, the contribution of $ \Gamma^\alpha_{[\mu \nu]}$ cannot be neglected and  spacetime can be described both by  curvature and  torsion
 \cite{Hehl}. 
 
For the sake of completeness, it is worth mentioning that the most general connection can be written even without  assuming metricity, i.e. $g_{\mu \nu; \, \alpha} = 0$, which, once relaxed, gives rise to a class of theories where   \emph{non-metricity} plays a crucial role in the description of  spacetime. In \cite{BeltranJimenez:2019tjy}, it is shown that actions constructed by   curvature,  torsion or  non-metricity scalars are  equivalent at the level of field equations since they differ only for  boundary terms. In this perspective, we can deal with the so-called  \emph{geometric trinity of gravity} \cite{BeltranJimenez:2019tjy}. 

The theory  involving only  the torsion scalar, with vanishing curvature in a purely affine formulation, is the   \emph{Teleparallel Equivalent of General Relativity} (TEGR); geometric foundations and applications of TEGR are discussed \emph{e.g.} in \cite{Arcos:2005ec, Cai:2015emx, Aldrovandi:2013wha}. It turns out that such a theory can be recast as a gauge theory for the translation group in a reference frame with vanishing spin connection.

Despite the need of extending/modifying GR,   self-interaction potentials, couplings and kinetic terms give rise to infinite choices which can lead to a frustrating indetermination  in fitting  observations and addressing conceptual problems. One can adjust models and parameters to match  single datasets and phenomena but a theory in agreement with the whole phenomenology seems far to be achieved. In other words, any single  theory loses its general  predictive power and cannot be used to reproduce a self-consistent  cosmic history starting from ultraviolet to infrared scales.  
Therefore,  some selection criteria, based on physical requirements, are needed to discriminate among the plethora of modified gravities.  These criteria can be based on symmetries, conservation laws and on general physical motivations.

Here, we want to consider scalar fields non-minimally coupled with different  geometric invariants, in particular the Ricci scalar $R$, the torsion scalar $T$, and the Gauss-Bonnet scalar $\G$. The aim is to demonstrate that all these non-minimally coupled invariants can give rise to similar cosmological dynamics once we know how to transform each-other. Furthermore, these scalar-tensor theories can be dealt under the standard of    \emph{Noether Symmetry Approach} which allows to fix  couplings, potentials and kinetic terms  requiring  the existence of symmetries and related conserved quantities. In other words, the  purpose  is to compare, by the Noether symmetries,  the dynamics of three different actions, containing a scalar field non-minimally coupled to different geometric invariants pointing out the equivalence  of the three representations of gravity.  The final result of this study is that theories admitting the same Noether symmetries has the same dynamics and the same solutions. In this sense, they are equivalent.

 The paper is organized as follows: in Sec. \ref{modified} we briefly argue some basic concepts about extended and  modified theories of gravity.  Sections \ref{Riccisect}, \ref{Torsionsect},  and \ref{Gausssect} are devoted to the discussion of non-minimal coupling with $R$, $T$, and $\G$ invariants through the Noether symmetries. We derive the form of couplings, kinetic term and potential according to the existence of symmetries. After, the reduction of related dynamical systems, cosmological solutions are found out and compared in the three cases. 
 In Sec.\ref{Hamiltoniandynamics}, thanks to the cyclic variables derived from the Noether method, we find the related Hamiltonians and compare  the three theories also at this level. 
 In  Sec. \ref{conc}, we draw our conclusions. Appendix \ref{noether} is a brief summary of the Noether Symmetry Approach.
 
\section{Extended and Modified Gravity}
\label{modified}
As already argued in the introduction, some of the shortcomings suffered by GR can be cured by extending the Einstein theory introducing in the action other curvature invariants like functions of $R$,  $R^{\mu\nu}R_{\mu\nu}$, $R^{\alpha\beta\mu\nu}R_{\alpha\beta\mu\nu}$, $W^{\alpha\beta\mu\nu}W_{\alpha\beta\mu\nu}$  where $R_{\mu\nu}$, $R_{\alpha\beta\mu\nu}$, and $W_{\alpha\beta\mu\nu}$ are the Ricci, Riemann, and Weyl tensors respectively, or considering  alternative  geometrical representations of the gravitational field like   the torsion scalar \cite{Cai:2015emx} or the non-metricity \cite{BeltranJimenez:2019tjy}. In the first case, we are dealing with \emph{Extended Gravity} \cite{Capozziello:2011et} because GR is a particular case of a  large class of models while, in the ore cases, we are dealing with \emph{Modified Gravity} because dynamics is not given by $R$ in the action. For a discussion see \cite{Capozziello:2019klx}.  Also if conceptual foundations of these approaches are different (e.g. based on Equivalence Principle, affinities, non-metricity), phenomenology which  they have to address is the same so the equivalence to GR and its extensions should be required for a self-consistent description of gravitational interaction.

An example of this statement can be given considering TEGR. 
The  torsion tensor $T^\alpha_{\mu \nu}$ is defined as
\begin{equation}
\Gamma^\alpha_{\,\,\,[\mu \nu]} := T^\alpha_{\,\,\,\mu \nu}\,.
\end{equation}
The most general connection involving curvature and torsion can be written as 
\begin{equation}
\Gamma^\alpha_{\,\,\, \mu \nu} = \breve{\Gamma}^\alpha_{\,\,\, \mu \nu}  + \frac{1}{2}\left(T^{\,\, \alpha}_{\mu \,\,\,\nu} + T_{\mu \nu}^{\,\,\,\alpha} - T^{\alpha}_{\,\, \mu \nu} \right),
\label{general connection}
\end{equation} 
being $\breve{\Gamma}^\alpha_{\,\,\, \mu \nu}$ the Levi-Civita connection. Furthermore, by defining the contorsion tensor as
\begin{equation}
K^\nu_{\;\;p \mu} = \frac{1}{2}(T_{p \;\; \mu}^{\;\;\nu} + T_{\mu \;\; p}^{\;\;\nu} - T^\nu_{\;\; p \mu})
\label{contorsione1}
\end{equation}
and the superpotential as
\begin{equation}
S^{p \mu \nu} = K^{\mu \nu p} - g^{p \nu} T^{\sigma \mu}_{\;\;\;\;\sigma} + g^{p \mu} T^{\sigma \nu}_{\;\;\;\;\sigma},
\label{superpotenziale}
\end{equation}
we derive the  \emph{torsion scalar}:
\begin{equation}
T = T_{p \mu \nu} S^{p \mu \nu}.
\label{definizione di torsione}
\end{equation}
Setting to zero the curvature, we obtain a theory of gravity, equivalent to GR,  described by  torsion.  Affine transformations are described by  the Weitzenb\"ock connection.  
As stated above, it turns out that TEGR can be seen as a gauge theory for translation group in the tangent spacetime, where the fundamental field is given by the \emph{tetrads}. Tetrad fields $e^a_\mu$ link the flat spacetime to the curved spacetime by means of the relation $g_{\mu \nu} = e^a_\mu e^b_\nu \eta_{ab}$, being $\eta_{ab}$ the Minkowski tensor. See \cite{Arcos:2005ec, Aldrovandi:2013wha,Abedi:2018lkr, Wei:2011yr, Ferraro:2006jd, Nashed:2014sea, Bahamonde:2016grb, Capozziello:2019klx} and reference therein.

However,
it turns out that GR  and TEGR  are linked through the relation $ \breve{R} = - T + B$, where $B$ is a boundary term \cite{Bahamonde:2016grb}. 
This relation has practical applications. For instance, in a spatially flat cosmological \emph{Friedmann-Lema\^itre-Robertson-Walker} (FLRW) metric of the form $g_{\mu \nu} = \text{diag} (1,-a(t)^2, -a(t)^2, -a(t)^2)$, the torsion takes the form
\begin{equation}
T = - 6 \left(\frac{\dot{a}}{a}\right)^2.
\label{Torsion cosmo}
\end{equation}
The Ricci scalar is
\begin{equation}
\label{Ricci}
R = - 6 \left(\frac{\ddot{a}}{a}+\frac{\dot{a}^2}{a^2}\right)\,,
\end{equation}
then immediately  the relation between $R$ and  $T$ can be recovered through the boundary term $B$:
\begin{equation}
B=- 6 \left( \frac{\ddot{a}}{a} + 2\frac{\dot{a}^2}{a^2} \right)\,.
\end{equation}
This means that the description of gravity of TEGR and GR, at cosmological level, is substantially equivalent. Differences emerge for $f(T)$ and $f(R)$ because the degrees of freedom of these two theories are different \cite{Cai:2015emx}. 

This is just an example of how two different representations of gravity can be practically connected. Below, we will discuss non-minimal coupling with $R$, $T$, and $\G$ scalars pointing out that analogue (or identical) features emerge if couplings, kinetic terms, and potentials are selected by Noether symmetries.

A general scalar-tensor action, written in terms of the scalar curvature, reads as:
\begin{equation}
S = \int \sqrt{-g} \left\{F(\phi) \; R  +\omega(\phi) g^{\mu \nu} \phi_{;\mu} \phi_{;\nu} - V(\phi) \right\} d^4x,
\label{action scalar field}
\end{equation}
where $F(\phi)$ is the coupling, $\omega(\phi)$ the coefficient of the kinetic term and $V(\phi)$ the potential. According to the previous considerations, such a theory can be both an extended gravity and a modified gravity depending on how GR, that is $R$ with minimal coupling is recovered, i.e. $F(\phi)\rightarrow F_0$, with $F_0$ the Newton constant and $\omega(\phi), V(\phi)\rightarrow 0$. The variation of the action with respect to  the metric tensor $g_{\mu \nu}$, provides the field equations  \cite{Capozziello:2014mea}
\begin{equation}
\label{field}
R_{\mu \nu} F(\phi) - \frac{1}{2} g_{\mu \nu} \left[R F(\phi) + \omega(\phi)  \phi_{; \alpha}\phi^{; \alpha} - V(\phi) \right] - F(\phi)_{; \mu ; \nu} + g_{\mu \nu} \Box F(\phi) + \omega(\phi) \phi_{; \mu} \phi_{; \nu} = 0\,,
\end{equation}
that clearly reduce to the Einstein equations considering the above conditions. Action \eqref{action scalar field} is the paradigm for a very large class of theories. For example,  $f(R)$ gravity in metric formalism can be easily recovered from  \eqref{action scalar field} if $\omega(\phi)$ is set to zero  and 
\begin{equation}
V(\phi) = \frac{1}{2} \frac{f(R) - R f_R}{[f_R]^2}\,,
\end{equation}
where the lower index means the derivative with respect to $R$. See \cite{Capozziello:2011et,Allemandi} for details.

In general, minimally coupled Einstein GR plus a scalar field can be recovered by a conformal transformation
\begin{equation}
\tilde{g}_{\mu \nu} = e^{2 \Omega} g_{\mu \nu}\,,
\end{equation}
where
\begin{eqnarray}
&& \tilde{\Gamma}^\sigma_{\lambda \mu} = \Gamma^\sigma_{\lambda \mu} + g^{\nu \sigma} \left(\partial_\lambda \Omega \; g_{\mu \nu} + \partial_\mu \Omega \; g_{\lambda \nu} - \partial_\nu \Omega \; g_{\lambda \mu} \right) \nonumber
\\
&& \tilde{R}_{\alpha \beta} = R_{\alpha \beta} - 2 \Omega_{; \alpha ; \beta} + 2 \Omega_{; \alpha} \Omega_{; \beta} - g_{\alpha \beta} \Box \Omega - 2 g_{\alpha \beta} \; \Omega_{; \gamma} \Omega^{; \gamma} \nonumber
\\
&& \tilde{R} = e^{-2 \Omega} \left(R - 6\Box  \Omega - 6 \Omega_{; \gamma} \Omega^{; \gamma} \right)\,,
\label{gamma conforme}
\end{eqnarray}
are the conformally transformed connection, Ricci tensor and Ricci scalar, respectively.
According to this transformation, the stress-energy tensor for the conformal field is defined as 
\begin{equation}
\tilde{T}_{\alpha \beta} = \phi_{; \alpha} \phi_{; \beta} - \frac{1}{2} \tilde{g}_{\alpha \beta} \phi_{; \gamma} \phi^{; \gamma} + \tilde{g}_{\alpha \beta} V(\phi)\,. 
\end{equation}
 Depending on the form of the functions, the scalar tensor action  provides several  features useful to describe  early and late time evolution of the universe. In particular, it can be   used in string-dilaton cosmology (see for example \cite{Nojiri:2006ri, Callan:1985ia, Damour:1994zq, Gasperini:2002bn}). 

In order to select the unknown functions $F(\phi), \omega(\phi)$ and $V(\phi)$, we can adopt the \emph{Noether Symmetry Approach}, summarized in  Appendix \ref{noether}. By Noether symmetries, non-minimal couplings with different geometric invariants can be dealt under the same standard.

\section{Non-Minimally Coupled Curvature Scalar}
\label{Riccisect}
\label{sectR}
Let us start our analysis considering action \eqref{action scalar field} where we will fix the form of unknown functions by Noether symmetries.   After selecting these  functions,  we will  find out  analytic cosmological solutions thanks to the reduction of  dynamics. 

From the variation of the action with respect to the scalar field $\phi$, we get the Klein-Gordon equation
\begin{equation}
\omega_\phi(\phi) \phi^{; \alpha} \phi_{; \alpha} + 2 \omega(\phi) \Box \phi - R F_\phi(\phi) + V_\phi(\phi) = 0,
\label{KG eq R}
\end{equation}
where the subscript $\phi$ denotes the derivative with respect to $\phi$. Together with Eqs.\eqref{field}, it completes the set of field equations.

The action can be simplified by focusing on a spatially-flat FLRW metric and integrating  over the 3-D surface term, so we get
\begin{equation}
S = 2 \pi^2 \int  a^3 \left[R F(\phi) + \omega(\phi) \dot{\phi}^2  - V(\phi) \right] dt.
\end{equation}
Finally, replacing the cosmological expression of the Ricci scalar \eqref{Ricci} into the action and integrating out second order derivatives, we get the cosmological point-like  Lagrangian
\begin{equation}
\Lagr = 6 F(\phi)a\dot{a}^2 + 6 F_\phi(\phi)a^2 \dot{a} \dot{\phi} + 
 a^3 \omega(\phi) \dot{\phi}^2 - a^3 V(\phi).
 \label{point-like R}
\end{equation} 
Field equations result in the Euler-Lagrange equations of \eqref{point-like R}. The energy condition $E_\Lagr = 0$, that is the $00$ equation, has to be also included for  consistency. Finally, we have the dynamical system:
\begin{equation}
\begin{cases}
a: \displaystyle 2 F \dot{a}^2 + 4 a \left(F_\phi \dot{a} + F \ddot{a}\right) + a^2 \left(V- \omega\dot{\phi}^2 + 2 F_{\phi \phi} \dot{\phi}^2 + 2 F_\phi \ddot{\phi} \right)=0
\\
\phi: \displaystyle 6 \dot{a}^2 F_\phi + 6 a \omega \dot{a} \dot{\phi} + a \left[6 F_\phi \ddot{a} + a \left(V_\phi + \dot{\phi}^2w_\phi +    2 w \ddot{\phi} \right)\right]=0
\\
E_\Lagr = 0: \displaystyle 6 F\dot{a}^2 + 6 a F_\phi \dot{\phi} \dot{a} + a^2 \left(V+ w \dot{\phi}^2\right)=0.
\end{cases}
\label{ELR}
\end{equation}
We notice that the system is completely equivalent to that coming from  the  field equations derived from the variation of the action with respect to the metric and the scalar fiend, once the FLRW metric is imposed. It can be solved after  the three functions of $\phi$ are selected  through Noether symmetries. The approach can be developed in the two-dimensional minisuperspace  $\mathcal{S} = \{a,\phi\}$ whose corresponding symmetry generator is
\begin{equation}
\mathcal{X} =\xi(a,\phi,t) \partial_t + \alpha(a,\phi,t) \partial_a + \beta(a,\phi,t) \partial_\phi\,.
\end{equation}
After equating to zero  terms containing same time derivatives of  variables, the application of Noether's identity \eqref{theorem2 cosmo} (see Appendix \ref{noether}) to Lagrangian \eqref{point-like R} provides a system of 10 differential equations. Nevertheless, by imposing \emph{a priori} the condition $\xi = \xi(t)$ (holding for Lagrangians in canonical forms) and neglecting redundant equations, the system reduces to 4 differential equations plus the condition on the infinitesimal generators, namely:
\begin{equation}
\begin{cases}
\displaystyle 3 \alpha + \beta V_\phi + V \partial_t \xi =0
\\
\displaystyle \beta \left[ 2 F_\phi V_\phi^2 + V^2 \omega_\phi - V \left( \omega V_\phi + 2 F_\phi V_{\phi \phi} \right) \right] - 2 V \left[V \omega \left( \partial_t \xi - \partial_\phi \beta \right) + F_\phi V_\phi \partial_\phi \beta \right] = 0
\\
\displaystyle \beta \left[2 F V_\phi^2 + 3 V^2 F_{\phi \phi} - V (3 F_\phi V_\phi +  2 F V_{\phi \phi})\right] + V^2 \left[F_\phi \left( - 6 \partial_t \xi + 3 \partial_\phi \beta\right) + a \omega \partial_a \beta \right] - V V_\phi \left( 2 F \partial_\phi \beta + a F_\phi \partial_a \beta \right) = 0
\\
\displaystyle 3 \beta \left(V F_\phi - F V_\phi \right) + 3 a V F_\phi \partial_a \beta -2 F \left(3 V \partial_t \xi +a V_\phi \partial_a \beta \right) = 0
\\
\displaystyle \alpha = \alpha(a,\phi) \quad \beta = \beta(a,\phi) \quad \xi = \xi(t).
\end{cases}
\end{equation}
The above system is clearly over determined and cannot provide any explicit form without imposing some constraint (see \cite{Capozziello:1996bi} for details). Since we want to investigate  functions with physical meaning for cosmology, we replace into the system both power-law and exponential potentials, so that it provides the following solutions:
\begin{equation}
\begin{cases}
\displaystyle \mathcal{X} = \left(\xi_0 t + \xi_1 \right) \partial_t - \frac{\xi_0}{3} a \left( \frac{k+c}{k-c} \right) \partial_a + \frac{2 \xi_0}{k-c} \phi \partial_\phi
\\
\displaystyle F(\phi) = F_0 \phi^k \quad \omega(\phi) = \omega_0 \phi^{k-2} \quad V(\phi) = V_0 \phi^c  \quad k \neq c
\\
\\
\displaystyle \mathcal{X} = \frac{k \alpha_0}{3} a^{- \frac{1}{5}} \partial_a +\alpha_0 a^{- \frac{6}{5}} \partial_\phi
\\
\displaystyle F(\phi) = F_0 \phi^k \quad \omega(\phi) = \omega_0 \phi^{k-2} \quad V(\phi) = V_0 \phi^k  \quad k = \pm \sqrt{\frac{2 \omega_0}{3 F_0}}
\\
\\
\displaystyle \mathcal{X} = \left(\xi_0 t + \xi_1 \right) \partial_t - \frac{\xi_0}{3} a \left( \frac{k+c}{k-c} \right) \partial_a + \frac{2 \xi_0}{k-c} \partial_\phi
\\
\displaystyle F(\phi) = F_0 e^{k \phi} \quad \omega(\phi) = \omega_0 e^{k \phi} \quad V(\phi) = V_0 e^{c \phi}  \quad k \neq c
\\
\\
\displaystyle \mathcal{X} = - \frac{k \alpha_0}{3}  a^{- \frac{1}{5}} e^{- \frac{k \omega_0 \phi}{F_0 k^2 + \omega_0}} \partial_a + \alpha_0 a^{-\frac{6}{5}} e^{- \frac{k \omega_0 \phi}{F_0 k^2 + \omega_0}}\partial_\phi
\\
\displaystyle F(\phi) = F_0 e^{k \phi} \quad \omega(\phi) = \omega_0 e^{k \phi} \quad V(\phi) = V_0 e^{k \phi}  \quad  k = \pm \sqrt{\frac{2 \omega_0}{3 F_0}}.
\end{cases}
\label{SoluzR}
\end{equation}
Furthermore, there is one further solution for constant coefficient of the kinetic term ($\omega(\phi)=1$), namely
\begin{equation}
  {\cal{X}} = - \frac{2(s+1)}{2s+3} \beta_0 a^{s+1} \phi^{\frac{2s^2+4s}{2s+3}} \partial_a + \beta_0 a^s \phi^{\frac{2s^2+6s+3}{2s+3}} \partial_\phi\,,
\qquad  F(\phi)  = \ell(s) \phi^2 \,,\quad V(\phi) = V_0 \phi^{\frac{6(s+1)}{2s+3}} \quad \ell(s) = \frac{(2s+3)^2}{48(s+1)(s+2)} ,
\label{second classR}
\end{equation}
with $\alpha_0, \beta_0, \xi_0, k,c,s, \omega_0, F_0,V_0$ real constants. We neglect trivial solutions, such as those with constant coupling or vanishing potential.  
Inserting the above functions into dynamics, it is reduced  and  equations of motion can be analytically solved. 

 Starting from the two main sets of functions selected above  we are going to obtain  exact cosmological solutions. It is worth noticing that the choice of exponential potential also leads to exponential coupling and exponential kinetic term like in  string-dilaton cosmology \cite{Capozziello:1993ts, Capozziello:1993vs, Capozziello:1993tr}. This means that the string-dilaton Lagrangian can be naturally obtained from Noether symmetries \cite{Capozziello:2015hra}. From this point of view, solutions occurring in Eq. \eqref{SoluzR} can be considered more general than those provided in \cite{Capozziello:1993vs}, since both the solutions outlined there by the authors are contained in the last two of Eq. \eqref{SoluzR}. In particular, the exponential potential of string-dilaton cosmology is recovered for $k=-2$ and arbitrary $c$, while the constant potential is recovered for for $k=-2$ and $c=0$. 

With these considerations in mind,  let us solve the Euler-Lagrange Eqs. \eqref{ELR} for   cases corresponding to the first and the third solution of Eq. \eqref{SoluzR}. Let us start with the former one. In this case,  the Lagrangian \eqref{point-like R} takes the form:
\begin{equation}
\Lagr = 6 F_0 \phi^k a\dot{a}^2 + 6 F_0 k \phi^{k-1} a^2 \dot{a} \dot{\phi} + 
 a^3 \omega_0 \phi^{k-2} \dot{\phi}^2 - a^3 V_0 \phi^c.
\end{equation} 
The Euler-Lagrange equations can be analytically solved  with the constraint $k = c = 2$ providing a de Sitter-like expansion of the form:
\begin{equation}
a(t) = a_0 e^{q t}\,, \quad \phi(t) = \phi_0 \exp\left\{\frac{1}{2}\left[ - 3q \pm \sqrt{\frac{-48 F_0 q^2 - 4 V_0 + 9 q^2 \omega_0}{\omega_0}} \right] t \right\}\,, \quad q, a_0, \phi_0 \in \mathbb{R}.
\label{cosmoR}
\end{equation}
Considering the third case of \eqref{SoluzR},  the point-like Lagrangian \eqref{point-like R} can be written as
\begin{equation}
\Lagr = 6 F_0 e^{k \phi} a\dot{a}^2 + 6 F_0 k e^{k \phi} a^2 \dot{a} \dot{\phi} + 
 a^3 \omega_0 e^{k \phi} \dot{\phi}^2 - a^3 V_0 e^{c \phi}\,,
\end{equation}
and, even in this case, the equations of motion set the value of the parameter $k$ and $c$, introducing the further constraint $k=c$. Therefore, discarding the solutions with minimal coupling, we find
\begin{equation}
a(t) = a_0 e^{qt} \,,\quad \phi(t) = \frac{3 F_0 k q \pm \sqrt{(3 F_0 k q)^2 - 6 F_0 \omega_0 q^2 - V_0 \omega_0}}{\omega_0}\,, \quad q \in \mathbb{R}.
\label{cosmoR1}
\end{equation}
The values of  the constants $F_0, \omega_0, V_0$ can be fixed according to cosmological  observations  \cite{Capozziello:2007iu}). In summary, deSitter-like expansions are provided by  symmetries. For the other cases, the analysis is similar.

 \section{Non-Minimally Coupled Torsion Scalar}
\label{Torsionsect}
\label{sectT}
With the above results in mind, let us  develop similar considerations for non-minimally coupled teleparallel gravity. We will show that  dynamics and solutions, derived from Noether symmetries,  are equivalent  to those obtained in  Sec. \ref{sectR}. In this sense, symmetries can be a criterion capable of comparing theories coming from different  representations of gravity. 

Let us consider the teleparallel equivalent of  action \eqref{action scalar field}, \emph{i.e.}:
\begin{equation}
S = \int e \left[T F(\phi) + \omega(\phi) \phi_{; \alpha} \phi^{; \alpha} - V(\phi) \right] d^4x,
\end{equation}
whose Klein-Gordon equation reads as
\begin{equation}
\omega_\phi(\phi) \phi^{; \alpha} \phi_{; \alpha} + 2 \omega(\phi) \Box \phi - T F_\phi(\phi) + V_\phi(\phi) = 0.
\end{equation}
Here $e$ takes the place of $\sqrt{-g}$ and stands for the determinant of tetrad fields. Unlike the previous case, the cosmological expression of  torsion does not contain  second derivatives which must be integrated out; therefore, the point-like Lagrangian can be easily found only by replacing the relation \eqref{Torsion cosmo} into the action and by integrating the three-dimensional surface:
\begin{equation}
\Lagr =  - 6 a F(\phi) \dot{a}^2 +  a^3 \omega(\phi) \dot{\phi}^2 -a^3 V(\phi).
\label{lagraT}
\end{equation}
Note that this  Lagrangian is already canonical and the equations of motion are simplified with respect to Eqs. \eqref{ELR}. They are:
\begin{equation}
\begin{cases}
a: \, -2 F \dot{a}^2 + a^2(V - \omega \dot{\phi}^2) - 4 a (F_\phi \dot{a} \dot{\phi}+ F \ddot{a}) = 0
\\
\phi: \, 6 \dot{a}^2 F_\phi + 6 a \omega \dot{a} \dot{\phi} + a^2 (V_\phi + \omega_\phi \dot{\phi}^2 +  2 \omega \ddot{\phi}) = 0
\\
E_\Lagr = 0: \, -6 a F \dot{a}^2 + a^3 (V + \omega \dot{\phi}^2) = 0.
\end{cases}
\label{ELT}
\end{equation}
In order to solve the system \eqref{ELT},  we  select  functions related to symmetries; the minisuperspace considered is two-dimensional as in the previous case ($\mathcal{S} = \{a,\phi\}$) and the  generator of the symmetry is, in turn,
\begin{equation}
\mathcal{X} =\xi(t) \partial_t + \alpha(a,\phi,t) \partial_a + \beta(a,\phi,t) \partial_\phi,
\end{equation}
where, being the Lagrangian in a canonical form, the condition $\xi = \xi(t)$ immediately holds. The application of the extended Noether vector to the point-like Lagrangian \eqref{lagraT} provides a system of 12 equations, which can be reduced to a system of 4 differential equations  with the constraints on the infinitesimal generators, that is:
\begin{equation}
\begin{cases}
6 F \partial_\phi \alpha  - 2 \omega a^2 \partial_a \beta = 0
\\
\alpha F + \beta a F_\phi + a F \partial_t \xi - 2 a F \partial_a \alpha = 0
\\
3 \alpha V + \beta a V_\phi - a V \partial_t \xi = 0
\\
3 \alpha \omega + \beta a \omega_\phi  - a \omega \partial_t \xi  + 2 a \omega \partial_\phi \beta = 0 
\\
\alpha = \alpha(a,\phi) \quad \beta = \beta(a, \phi) \quad \xi = \xi(t).
\end{cases}
\end{equation}
After some manipulations, the system can be recast as a system of two differential equations containing the three functions $F(\phi), \omega(\phi), V(\phi)$ and two unknown infinitesimal generators. It is therefore clear that the system cannot provide a unique solution, and an initial choice must be performed in order to fix the related  dynamics. However, the assumption is not too much strict, since only the form of the potential is needed in order to exactly solve the system. Solutions containing power-law and exponential potentials are:
\begin{equation}
\begin{cases}
\displaystyle \mathcal{X} = \left(\xi_0 t + \xi_1 \right) \partial_t - \frac{\xi_0}{3} a \left( \frac{k+c}{k-c} \right) \partial_a + \frac{2 \xi_0}{k-c} \phi \partial_\phi
\\
\displaystyle F(\phi) = F_0 \phi^k \quad \omega(\phi) = \omega_0 \phi^{k-2} \quad V(\phi) = V_0 \phi^c  \quad k \neq c
\\
\\
\displaystyle \mathcal{X} = \left(\xi_0 t + \xi_1 \right) \partial_t - \frac{\xi_0}{3} a \left( \frac{k+c}{k-c} \right) \partial_a + \frac{2 \xi_0}{k-c} \partial_\phi
\\
\displaystyle F(\phi) = F_0 e^{k \phi} \quad \omega(\phi) = \omega_0 e^{k \phi} \quad V(\phi) = V_0 e^{c \phi}   \quad k \neq c
\\
\\
\displaystyle \mathcal{X} = -\frac{k}{3} a \beta(\phi) \partial_a + \beta(\phi) \partial_\phi
\\
\displaystyle F(\phi) = F_0 e^{k \phi} \quad \omega(\phi) = \omega_0 e^{k \phi} \quad V(\phi) = V_0 e^{k \phi}  
\end{cases}
\label{SoluzT}
\end{equation}
and those with unitary kinetic term are
\begin{equation}
\begin{cases}
\displaystyle {\cal{X}} = - \frac{2 \beta_0}{2 s + 3} a^{s+1} \phi^{-\frac{2 s}{2 s +3}} \partial_a + \beta_0 a^s \phi^{\frac{3}{2 s +3}} \partial_\phi
\\
\displaystyle F(\phi) = \frac{(2 s + 3)^2}{48} \phi^2 \,\,\,\,\,\,\, V(\phi) = V_0 \phi^{\frac{6}{2 s + 3}}
\\
\\
\displaystyle {\cal{X}} = -\frac{2}{3} a^{\frac{1}{4}} (c_2 + 2 c_3 \phi) \partial_a +  a^{-\frac{3}{4}}(c_1 + c_2 \phi + c_3 \phi^2) \partial_\phi 
\\
\displaystyle F(\phi) = \frac{3}{64 c_3} (c_1 + c_2 \phi + c_3 \phi^2) \,\,\,\,\,\,\,\,\, V(\phi) = V_0(c_1 + c_2 \phi + c_3 \phi^2)^2.
\end{cases}
\label{second classT}
\end{equation}
Also in this case, the exponential solutions of Noether system allow us to find out the teleparallel equivalent of string-dilaton cosmology, namely the string-dilaton action with torsion instead of curvature. Once finding the functions, it is possible to get the exact cosmological solutions.

Let us now solve the Euler-Lagrange equations \eqref{ELT} for two different set of couplings, potentials and kinetic terms. We choose the most general solutions among those in \eqref{SoluzT}, namely the first and the second. In the former case the Lagrangian \eqref{lagraT} turns out to be:
\begin{equation}
\Lagr =  - 6 F_0 a \phi^k \dot{a}^2 + \omega_0 a^3 \phi^{k-2} \dot{\phi}^2 - V_0 a^3 \phi^c\,.
\end{equation}
Assuming the condition $k=c$ and the de Sitter-like expansion for the scale factor, we have:
\begin{equation}
a(t) = a_0 e^{q t}\,, \quad \phi(t) = \phi_0 \exp \left\{ \pm \sqrt{\frac{6 F_0 q^2 - V_0}{\omega_0}} \, t\right\}\,, \quad q = \sqrt{\frac{V_0 \omega_0}{6 F_0 \omega_0 - 4 k^2 F_0^2}}\,.
\label{cosmoT}
\end{equation}
By taking into account the second set of functions, the Lagrangian takes the form
\begin{equation}
\Lagr =  - 6 F_0 a e^{k \phi} \dot{a}^2 + \omega_0 a^3 e^{k \phi} \dot{\phi}^2 - V_0 a^3 e^{c \phi},
\end{equation}
leading to the exponential solutions constrained by the relation $k = c$:
\begin{equation}
a(t) = a_0 e^{q t}\,, \quad \phi(t) =  \pm \sqrt{\frac{6 F_0 q^2 - V_0}{\omega_0}} \, t\,, \quad q = \sqrt{\frac{V_0 \omega_0}{6 F_0 \omega_0 - 4 k^2 F_0^2}}\,.
\end{equation}
It is worth stressing  the difference between the scalar field coupled to the  curvature scalar and the corresponding torsion one. The Noether approach performed in Sec. \ref{sectR} allows to find exact expressions for the scalar field and for the scale factor, but the analytic relations between the free parameters cannot be obtained analytically. In the case treated here, instead, such a relation can be analytically found, so that an exact solution of Euler-Lagrange equations \eqref{ELT} occurs. This is due to the cosmological expression of $T$ which, not containing second derivatives, leads immediately to a canonical  Lagrangian. 

Similar results occurs considering the Gauss--Bonnet topological  term non minimally coupled to a scalar field, as we are going to discuss in the forthcoming section.

\section{Non-Minimally Coupled Gauss--Bonnet Scalar}
\label{Gausssect}

 Among extended theories of gravity, there is a particular combination of second-order curvature invariants which turns into a topological surface term which is  the Gauss--Bonnet topological scalar:  
\begin{equation}
\G = R^2 - 4 R^{\mu \nu}R_{\mu \nu} + R^{\mu \nu p \sigma}R_{\mu \nu p \sigma}\,.
\end{equation} 
It represents the four-dimensional Euler class, whose integral over the manifold provides the Euler Characteristic, according  to the \emph{Generalized Gauss--Bonnet Theorem} \cite{Fine:2012ze, Chern:1999jn, GBtheorem}. The Gauss--Bonnet scalar is a topological boundary term even in less than four dimensions and it is  non-trivial starting from  five dimensions. It naturally arises, for instance, in the Lovelock four-dimensional Lagrangian \cite{Easson:2020mpq} or in the Chern-Simons Lagrangian \citep{Zanelli:2015pxa}. Nevertheless, a non-linear function of the Gauss--Bonnet scalar provides non-trivial contributions to the field equations even in four dimensions, while starts being a surface term in less than four; for this reason it is possible to get non-trivial equations of motion from the action $S = \int \sqrt{-g} f(\G) d^4 x$, (with $f(\G) \neq \G$). In several papers,  the action containing the function $f(\G)$ is considered \cite{Saltas:2010ga, Rashidi:2018lwq} and it turns out that it provides a possible explanation of inflation \cite{Zhong:2018tqn,Paolella}, Black Hole solutions \cite{S.Silva:2018irj} or generalizations of the $\Lambda$CDM model \cite{Myrzakulov:2010gt}. Other works consider  functions depending on both  the Gauss--Bonnet term and  the Ricci scalar, obtaining  results in cosmology \cite{Li:2007jm}, alternatives to Dark Energy \cite{Nojiri:2005jg}, corrections to $\Lambda$CDM model \cite{Elizalde:2010jx} and other achievements. Most of them add into the action also the scalar curvature $R$; even though this choice is  useful  to recover GR as a limit,  the introduction of $R$ into the action is not the only way to recover GR, since the same can be obtained even considering a particular forms of  function $f(\G)$. As a matter of fact, it turns out that, in cosmology, the curvature invariants $R^{\mu \nu} R_{\mu \nu}$ and $R^{\mu \nu p \sigma} R_{\mu \nu p \sigma}$ are comparable with the term $R^2$. According to this statement, $R = C_0 \sqrt{|\G|}$, with $C_0$ being a constant, can be consistently adopted in cosmology\footnote{In the following, we are assuming the modulus of $\G$ when dealing with the square root because the invariant has to be real.}
(see \cite{baj2}). Similar analogies shows that a function of $\G$ also provides interesting solutions in  spherical symmetry \cite{baj}. Therefore,  non-linear functions of $\G$ can describe, in principle, phenomenology   at any scales. 

Here, we will consider functions of $\G$ non-minimally coupled with a scalar field in order to discuss solutions analogue to the above non-minimally coupled curvature and torsion cases.
Let us start considering the action
\begin{equation}
S = \int \sqrt{-g} \left[\G^n F(\phi) + \omega(\phi)  \phi_{;\alpha} \phi^{;\alpha} - V(\phi) \right] d^4 x\,, \quad \mbox{with} \quad n \in \mathbb{R},
\label{action GB}
\end{equation}
whose Klein-Gordon equation and field equations read respectively
\begin{equation}
\omega_\phi(\phi) \phi^{; \alpha} \phi_{; \alpha} + 2 \omega(\phi) \Box \phi - \G^n F_\phi(\phi) + V_\phi(\phi) = 0,
\label{KG eq G}
\end{equation}
\begin{align}\label{feq}
&\frac{1}{2} g_{\mu \nu} \G^n F(\phi)  - 2 n F(\phi)  \left(R R_{\mu \nu} - 2 R_{\mu \alpha} R^{\alpha}{}_\nu + R_\mu {}^{\alpha \beta \gamma} R_{\nu \alpha \beta \gamma} - 2 R^{\alpha \beta} R_{\mu \alpha \nu \beta}\right)  \G^{n-1}  +\nonumber
\\
&+ n F(\phi)  \left[2R \nabla_\mu \nabla_\nu +4 G_{\mu \nu} \Box - 4 (R^{\rho}{}_{ \nu} \nabla_{\mu} +R^{\rho}{}_{ \mu} \nabla_{\nu}) \nabla_{\rho} + 4 g_{\mu \nu} R^{\rho \sigma} \nabla_\rho \nabla_\sigma - 4 R_{\mu \alpha \nu \beta} \nabla^\alpha \nabla^\beta \right] \G^{n-1} + \nonumber
\\
&- \frac{1}{2} g_{\mu \nu} \omega(\phi)  \phi_{; \alpha}\phi^{; \alpha} - F(\phi)_{; \mu ; \nu} + g_{\mu \nu} \Box F(\phi) + \omega(\phi) \phi_{; \mu} \phi_{; \nu} + \frac{1}{2} g_{\mu \nu} V(\phi) = 0,
\end{align}
where $\nabla$ denotes the covariant derivative and, as above,  $\Box$ is the D'Alembert operator defined as $\Box = \nabla_\alpha \nabla^\alpha$.
Note that, with respect to Secs. \ref{sectR} and \ref{sectT}, we introduced into the action a new degree of freedom, hence the minisuperspace  in no longer two-dimensional, but it contains one more variable, that is  $\mathcal{S} = \{a, \phi, \G\}$. This is linked to the term $\G^n$ which cannot be treated at the same level of  $R$ and $T$ due to the power $n$. By replacing the cosmological expression of $\G$ into the action, we obtain second order derivatives which cannot be eliminated through a simple integration. In order to find out the point-like Lagrangian, we have to define a further  Lagrange multiplier and introduce into the action a new  Lagrange parameter $\lambda$ with the constraint 
\begin{equation}
\G =  24 \frac{\dot{a}^2 \ddot{a}}{a^3}. 
\end{equation}
After integrating the surface term, the action  turns out to be:
\begin{equation}
S = 2 \pi^2 \int a^3 \left\{\left[F(\phi) \G^n + \omega(\phi) \dot{\phi}^2 - V(\phi) \right] - \lambda \left( \G - 24 \frac{\dot{a}^2 \ddot{a}}{a^3} \right) \right\} dt.
\end{equation}
The Lagrange multiplier can be found by varying the action with respect to the Gauss--Bonnet invariant. It is:
\begin{equation}
\frac{\delta S}{\delta \G} = 0 \quad \to \quad \lambda = a^3 n \G^{n-1} F(\phi).
\end{equation}
Replacing now the result into the action and integrating out the second derivatives, the point-like Lagrangian becomes
\begin{equation}
\Lagr = (1 - n) a^3 \G^n F(\phi) - 8 n \dot{a}^3 \dot{\phi} F_\phi(\phi) \G^{n-1} + a^3\omega(\phi) \dot{\phi}^2 - a^3 V(\phi) - 8 n (n - 1) G^{n - 2} \dot{a}^3 \dot{\G} F(\phi).
\label{lagraG}
\end{equation}
Clearly the Gauss-Bonnet contribution disappears for $n=1$.
In this case we have three equations of motion and the energy condition; the further equation is the one for $\G$, which  provides the cosmological expression of the Gauss--Bonnet surface term by construction. The Euler-Lagrange equations therefore read
\begin{equation}
\begin{cases}
\displaystyle a: \, 3 a^2 \left[(n-1) F \G^n + V  - \omega \dot{\phi}^2\right] - 
 24 n G[t]^{n-3} \dot{a} \left\{ (n-1) F \left[2 \G \dot{\G} \ddot{a} + (n-2) \dot{a} \dot{\G}^2 + \dot{a} \G \ddot{\G}\right] + \right.
 \\
\left. + 2 \G^2 F_\phi \dot{\phi} \ddot{a} + \dot{a} \G \left[\G \dot{\phi}^2 F_{\phi \phi} + 2 F_\phi  (n-1) \dot{\G} \dot{\phi}+ \G F_\phi \ddot{\phi} \right] \right\} = 0
\\
\displaystyle \phi: \, 6 a^2 \omega \dot{a} \dot{\phi}- 24 n \G^{n-1} \dot{a}^2 F_\phi \ddot{a} + a^3 \left[(n-1) \G^n F_\phi + V_\phi + \omega_\phi \dot{\phi}^2 + 2 \omega \ddot{\phi} \right] = 0
\\
\displaystyle \G: \, a^3 \G - 24 \dot{a}^2 \ddot{a} = 0
\\
\displaystyle E_\Lagr = 0: \, -8 n \G^{n-2}\dot{a}^3 \left[2 (n-1)F \dot{\G} + 3 \G F_\phi \dot{\phi} \right]+ a^3 \left[(n-1) F \G^n + V + \omega \dot{\phi}^2 \right]= 0.
\end{cases}
\label{ELG}
\end{equation}
The generator of  symmetry in  three-dimensional minisuperspace contains one further infinitesimal generator related to the   $\G$:
\begin{equation}
\mathcal{X} =\xi(a,\phi,\G, t) \partial_t + \alpha(a,\phi,\G, t)  \partial_a + \beta(a,\phi,\G, t) \partial_\phi + \gamma(t,a,\phi,\G) \partial_\G
\end{equation}
so that the application of $X^{[1]}$ to Lagrangian \eqref{lagraG} provides the following system of 4 differential equations
\begin{equation}
\begin{cases}
(n-1) \left(\gamma F_\phi + F \partial_\phi \gamma \right) + \beta \G F_{\phi \phi} + F_\phi \G \left(- 3 \partial_t \xi + \partial_\phi \beta + 3 \partial_a \alpha \right) = 0
\\
\beta \G F_\phi + F \left[(n-2) \gamma + \G \left(-3 \partial_t \xi + \partial_\G \gamma + 3 \partial_a \alpha \right) \right] = 0
\\
3 \alpha \omega+ a \left[\beta \omega_\phi -  \omega (\partial_t \xi - 2 \partial_\phi \beta)\right] = 0
\\
3 \alpha \G \left[(n-1) F \G^n + V \right] +  a (n-1) F \G^n \left(n \gamma + \G \partial_t \xi\right) + a \G \beta \left[(n-1 ) \G^n F_\phi + V_\phi \right] + a \G V \partial_t \xi = 0
\\
\alpha = \alpha(a) \quad \beta = \beta(\phi) \quad \gamma = \gamma(a, \phi, \G) \quad \xi = \xi(t).
\end{cases}
\end{equation}
As in the previous cases, the system is overdetermined and admits an infinite class of solutions depending on the form of the unknown coupling, namely
\begin{eqnarray}
&&\alpha = \alpha_0 a \quad \beta = -\frac{(3 \alpha_0 + \xi_0 - 4 n \xi_0 )F (\phi)}{F_\phi(\phi)} \quad \gamma = -4 \xi_0 \G \quad \xi = \xi_0 t + \xi_1 \nonumber
\\
&& \omega = \frac{F(\phi)^{\frac{-3 \alpha_0 + (8n-3) \xi_0}{3 \alpha_0 + \xi_0 - 4 n \xi_0}}F_\phi(\phi)^2}{(3 \alpha_0 + \xi_0 - 4 n \xi_0)^2} \qquad \qquad \quad \,\,\,\, V = V_0 F(\phi)^{\frac{3 \alpha_0 + \xi_0}{3 \alpha_0 + \xi_0 - 4 n \xi_0}} .
\label{soluzG}
\end{eqnarray}
Therefore, by choosing exponential and power-law couplings, Eq. \eqref{soluzG} can be split in two different solutions:
\begin{eqnarray}
&&\alpha = \alpha_0 a \quad \beta = -\frac{3 \alpha_0 + \xi_0 - 4 n \xi_0}{k} \quad \gamma = -4 \xi_0 \G \quad \xi = \xi_0 t + \xi_1 \nonumber
\\
&& \omega = \frac{k^2}{(3 \alpha_0 + \xi_0 - 4 n \xi_0)^2} e^{\frac{k(3 \alpha_0 - \xi_0) \phi}{3 \alpha_0 + \xi_0 - 4 n \xi_0}} \qquad  V = V_0  F_0^{\frac{3 \alpha_0 + \xi_0}{3 \alpha_0 + \xi_0 - 4 n \xi_0}} e^{\frac{k(3 \alpha_0 + \xi_0) \phi}{3 \alpha_0 + \xi_0 - 4 n \xi_0}} \qquad F(\phi) = F_0 e^{k \phi} \label{noethG}
\\ \nonumber
\\  \nonumber
\\
&&\alpha = \alpha_0 a \quad \beta = -\left(\frac{3 \alpha_0 + \xi_0 - 4 n \xi_0}{k}\right) \phi \quad \gamma = -4 \xi_0 \G \quad \xi = \xi_0 t + \xi_1 \nonumber
\\
&& \omega = \frac{k^2}{(3 \alpha_0 + \xi_0 - 4 n \xi_0)^2} \phi^{\frac{k(3 \alpha_0 - \xi_0)}{3 \alpha_0 + \xi_0 - 4 n \xi_0}-2} \qquad  V = V_0  F_0^{\frac{3 \alpha_0 + \xi_0}{3 \alpha_0 + \xi_0 - 4 n \xi_0}} \phi^{\frac{k (3 \alpha_0 + \xi_0)}{3 \alpha_0 + \xi_0 - 4 n \xi_0}} \qquad F(\phi) = F_0 \phi^k.
\label{NoethersolG}
\end{eqnarray}
The above hold as long as $n \neq 1$; otherwise we obtain the following:
\begin{eqnarray}
&&\alpha = 0 \quad \beta = \frac{3 \xi_0}{k} \phi \quad \xi = \xi_0 t + \xi_1 \quad F(\phi) = - \frac{1}{3 V_0^2} \phi^k \quad V(\phi) = V_0 \phi^{-\frac{k}{3}} \quad \omega(\phi) = \omega_0 \phi^{\frac{k}{3}-2}
\\
&& \alpha = 0 \quad \beta = \frac{3 \xi_0}{k} \phi \quad \xi = \xi_0 t + \xi_1 \quad F(\phi) = \frac{1}{k} e^{k \phi} \quad V(\phi) = V_0 e^{-\frac{k}{3} \phi} \quad \omega(\phi) = \omega_0 e^{\frac{k}{3}\phi}
\end{eqnarray}
where, as before, $\xi_0, \alpha_0, \beta_0, \gamma_0, k, n$ are real constants. The $n=1$ limit is topologically trivial only when $k = 1$, where the  contribution of the geometry in the corresponding action turns into a topological surface term. For this reason, there is no interest in investigating  cosmological solutions occurring for $k = n = 1$ and, in what follows, we will only focus on the $n=1/2$ case, which represents the Gauss--Bonnet equivalent to GR in the cosmological framework.

Let us derive cosmological solutions for  the Noether symmetry  \eqref{soluzG}. We will solve the Euler-Lagrange equations \eqref{ELG}  for $n = 1/2$ in order to compare the results with the above curvature case.  For $f(\G)= \G^{1/2}$, Noether's solutions \eqref{NoethersolG} can be written as:
\begin{eqnarray}
&&\alpha = \frac{\ell}{6}(z+k) a \quad \beta = - \ell \quad \gamma = -2 \ell(z-k) \G \quad \xi = \frac{\ell}{2} (z-k) t + \xi_1 \nonumber
\\
&& \omega(\phi)  = \frac{1}{\ell^2} e^{k \phi} \qquad  V(\phi)  = \tilde{V}_0  e^{z \phi} \qquad F(\phi) = F_0 e^{k \phi} \label{NoethersolG2}
\\ \nonumber
\\
&&\alpha = \frac{\ell}{6}(z+k) a \quad \beta = - \ell \phi \quad \gamma = -2 \ell(z-k) \G \quad \xi = \frac{\ell}{2} (z-k) t + \xi_1 \nonumber
\\
&& \omega(\phi)  = \frac{1}{\ell^2} \phi^{k-2} \qquad  V(\phi)  = \tilde{V}_0 \phi^{z} \qquad F(\phi) = F_0 \phi^k,
\label{NoethersolG1}
\end{eqnarray}
where we have defined 
\begin{equation}
\ell \equiv \displaystyle \frac{3 \alpha_0 - \xi_0}{k} \qquad V_0   F_0^{1+\frac{2 \xi_0}{k \ell}} \equiv \tilde{V}_0 \qquad z \equiv k \left(\frac{3 \alpha_0 + \xi_0}{3 \alpha_0 - \xi_0} \right)\,.
\end{equation} 
Let us start by analyzing the action containing a power-law coupling, namely:
\begin{equation}
S = \int \sqrt{-g} \left[F_0 \sqrt{\G} \phi^k + \frac{1}{\ell^2}\dot{\phi}^2 \phi^{k-2} + \tilde{V}_0 \phi^z \right] d^4 x,
\end{equation}
corresponding to the solution of Eq. \eqref{NoethersolG}. After solving the system \eqref{ELG}, we  obtain a de Sitter-like solution which fixes the values of $k$ and $z$ to $\displaystyle k = z = \frac{\sqrt{6}}{2 \ell^2 F_0}$. It reads as:
\begin{eqnarray}
a(t) = a_0 e^{- \frac{\sqrt{- 6 \tilde{V}_0}}{2 \ell F_0} t}\,, \quad \phi(t) = \phi_0 e^{\ell \sqrt{- \tilde{V}_0} t}\,,  \quad \G(t) = \frac{54 \tilde{V}_0^2}{\ell^4 F_0^4}. 
\label{cosmoG}
\end{eqnarray}
This means that, by merging the result provided by the Euler-Lagrange equations with those coming from the Noether approach, the only  generator associated to this case is:
\begin{equation}
\mathcal{X} = \frac{1}{\sqrt{6} \ell F_0} a \partial_a - \ell \phi \partial_\phi,
\end{equation}
which describes an internal gauge symmetry. Let us now analyze the second solution with exponential coupling, potential and kinetic term; the corresponding action takes the form:
\begin{equation}
S = \int \sqrt{-g} \left[F_0 \sqrt{\G} e^{k \phi} + \frac{1}{\ell^2} \dot{\phi}^2 e^{k \phi} + \tilde{V}_0 e^{z \phi} \right] d^4 x.
\end{equation}
By replacing Eq. \eqref{NoethersolG2} into the equations of motion \eqref{ELG}, it turns out that these latter can be analytically solved  by imposing the constraint $ k = z$, so that the scale factor and the scalar field behave like
\begin{equation}
a(t) = a_0 \exp \left\{\frac{k \ell}{3} \sqrt{\tilde{V}_0} (1+\sqrt{2}) \, t\right\}\,, \qquad \phi(t) = - \ell \sqrt{\tilde{V}_0 } t \,,\qquad \G(t) = \frac{8 k^4 \ell^4}{27} \tilde{V}_0^2 (1 + \sqrt{2})^4\,, \qquad k = \frac{3}{\ell^2} \sqrt{\frac{1}{F_0 \sqrt{21 - 12 \sqrt{2}}}}.
\end{equation}
As final remark, it is worth noticing that the non-minimal couplings with the invariants $R$, $T$, and $\G$ all admit de Sitter solutions which can be easily compared each-other. It is important to point out that the Gauss-Bonnet topological  invariant can be defined also in the case of teleparallel gravity \cite{Manos,Kostas} so that the above representations of gravity can be made totally equivalent also at this level.

\section{Equivalence of Hamiltonian Dynamics}
\label{Hamiltoniandynamics}
We want to show now that for internal symmetries, namely for $\xi(t) = 0$, the Noether Approach provides  transformation laws allowing to introduce  cyclic variables into the point-like cosmological Lagrangian. In order to get internal symmetries, we have to set the infinitesimal generator, related to the time variation, equal to zero. In the case of Ricci scalar coupled to the scalar field, the only solution containing symmetries which, after setting $\xi(t) = 0$, does not lead to trivial results, is that written in Eq. \eqref{second classR} (see also \cite{Capozziello:1996bi}). With regards to the torsional case, the only compatible solutions are \eqref{second classT}, while, for the $\sqrt{\G}$ scalar, both solutions can be equivalently considered. It is worth noticing that the generators in \eqref{second classR}, the first in Eq. \eqref{second classT} and that in Eq. \eqref{NoethersolG1}, under appropriate conditions, are  equivalent. For this reason, only the Hamiltonian dynamics provided by the following generators will be investigated:
\begin{equation}
\qquad \qquad \qquad \qquad \quad \,\, \,\, R: \to \begin{cases}
\displaystyle  {\cal{X}} = - \frac{2(s+1)}{2s+3} \beta_0 a^{s+1} \phi^{\frac{2s^2+4s}{2s+3}} \partial_a + \beta_0 a^s \phi^{\frac{2s^2+6s+3}{2s+3}} \partial_\phi
\\
\displaystyle  F(\phi)  = \frac{(2s+3)^2}{48(s+1)(s+2)} \phi^2 \,\,\,\,\, V(\phi) = V_0 \phi^{\frac{6(s+1)}{2s+3}},
\end{cases}
\label{solRxi0}
\end{equation} 
\begin{equation}
\qquad \qquad \qquad \,\, T: \to \begin{cases}
\displaystyle {\cal{X}} = - \frac{2 \beta_0}{2 s + 3} a^{s+1} \phi^{-\frac{2 s}{2 s +3}} \partial_a + \beta_0 a^s \phi^{\frac{3}{2 s +3}} \partial_\phi
\\
\displaystyle F(\phi) = \frac{(2 s + 3)^2}{48} \phi^2 \,\,\,\,\,\,\, V(\phi) = V_0 \phi^{\frac{6}{2 s + 3}}
\end{cases}
\label{solTxi0}
\end{equation} 
\begin{equation}
\qquad \qquad \qquad \qquad \G: \to \begin{cases}
\displaystyle \mathcal{X} = \frac{k \ell}{3} a \partial_a - \ell \phi \partial_\phi 
\\
\displaystyle F(\phi) = F_0 \phi^k \qquad V = V_0 \phi^{k} \qquad \omega(\phi) = \frac{1}{\ell^2} \phi^{k-2}\,.
\end{cases}
\label{solGxi0}
\end{equation} 
In the last solution,  to obtain the condition $\xi = 0$, we set $k = z, \, \xi_1 = 0$. In order to  compare the solutions, we set the coefficient of the kinetic term in the Gauss--Bonnet case \eqref{solGxi0} to be constant, as  naturally provided by the Noether Approach in the other two cases \eqref{solRxi0} and \eqref{solTxi0}. Therefore, by setting $s = 0$ and $k=2$, \eqref{solRxi0}, \eqref{solTxi0} and \eqref{solGxi0} become
\begin{equation}
R: \to \begin{cases}
\displaystyle  {\cal{X}} = - \frac{2}{3} \beta_0 a \partial_a + \beta_0 \phi \partial_\phi
\\
\displaystyle  F(\phi)  = \frac{3}{32} \phi^2 \,\,\,\,\, V(\phi) = V_0 \phi^2 ,
\end{cases}
\end{equation} 
\begin{equation}
T: \to \begin{cases}
\displaystyle {\cal{X}} = - \frac{2 \beta_0}{3} a \partial_a + \beta_0 \phi \partial_\phi
\\
\displaystyle F(\phi) = \frac{3}{16} \phi^2 \,\,\,\,\,\,\, V(\phi) = V_0 \phi^2
\end{cases}
\end{equation} 
\begin{equation}
\G: \to \begin{cases}
\displaystyle \mathcal{X} = \frac{2 \ell}{3} a \partial_a - \ell \phi \partial_\phi 
\\
\displaystyle F(\phi) = F_0 \phi^2 \qquad V = V_0 \phi^{2} 
\end{cases}
\end{equation} 
that is the symmetries fix\footnote{In the case of the Gauss-Bonnet coupling, the vector $\mathcal{X}$ is equivalent to the previous cases multiplying by -1. } the equivalence among the three representations of gravity when a scalar field is coupled with $R$, $T$, and $\sqrt{\G}$. This is the main result of this paper.

 To finalize our approach, let us consider  the generator \eqref{solRxi0}.  Thanks to the system \eqref{inner} in the Appendix \ref{noether}, we can perform the change of variables induced by the  Noether symmetry which allows to introduce a cyclic variable into the Lagrangian \cite{Capozziello:1996bi}.   The system \eqref{inner} takes the form:
\begin{equation}
\begin{cases}
\displaystyle \mathcal{X} z = \frac{2 \beta_0}{3} a \partial_a z - \beta_0 \phi\partial_\phi z = 1
\\
\displaystyle \mathcal{X} u = \frac{2}{3} a \partial_a  u - \phi \partial_\phi u = 0,
\label{sist1}
\end{cases}
\end{equation}
where $z$ represents the cyclic variable and the minisuperspace of configurations is trasformed from $\mathcal{S} = \{a,\phi \}$ to $\mathcal{S}' = \{z,u\}$. A possible solution of the above system is
\begin{equation}
\begin{cases}
\displaystyle z = - \frac{1}{\beta_0} \ln \phi
\\
\displaystyle u = a^{\frac{3}{2}} \phi
\end{cases}
\qquad \to \qquad
\begin{cases}
\displaystyle \phi = e^{-\beta_0 z}
\\
\displaystyle a = u^\frac{2}{3} e^{\frac{2 \beta_0}{3} z}.
\end{cases}
\label{olvarR}
\end{equation}
Replacing  the new variables $u,z$ into the Lagrangian \eqref{point-like R},  we get 
\begin{equation}
\Lagr_R =  - V_0 u^2 + \frac{1}{2} \ell^2 u^2\dot{z}^2 - \frac{1}{4} \ell u \dot{u} \dot{z} + \frac{1}{4} \dot{u}^2,
\end{equation}
where we set $\ell \equiv \beta_0$ in order to conform the notation to the other examples. Clearly this form of $\Lagr_R$ is cyclic in $z$. After finding the time-derivatives of the variables as functions of the conjugate momenta, we can easily get the Hamiltonian:
\begin{equation}
\mathcal{H}_R = \pi_u^2 + \frac{4}{7 \ell^2} \frac{\pi_z^2}{u^2} + V_0 u^2.
\end{equation}
Classical trajectories \eqref{cosmoR} can be recovered by means of the Hamilton--Jacobi equations by going back to the old variables \eqref{olvarR}. It is worth noticing  that, in order to provide a comparison among the three equivalent cases, the Hamiltonian dynamics has been studied for the solution \eqref{solRxi0} only; however, the change of variables coming from the Noether approach can be also found for the other solutions of Noether system, as shown, \emph{e.g.}, in \cite{Capozziello:1994ie, Capozziello:2013qha, Capozziello:1994du, Paliathanasis:2014rja, Capozziello:1994dn, Capozziello:1996bi}. 

The next case  is the torsion non-minimally coupled to the scalar field. Let us  focus on the solution \eqref{solTxi0} with $s = 0$. With this assumption,  the system providing the suitable  change of variables is:
\begin{equation}
\begin{cases}
\displaystyle -\frac{2 \beta_0}{3}a \partial_a z + \beta_0 \phi  \partial_\phi z = 1
\\
\\
\displaystyle -\frac{2}{3} a \partial_a u + \phi \partial_\phi u = 0 \;,
\end{cases}
\label{sist2}
\end{equation}
whose possible solution is the same as before, namely
\begin{equation}
\begin{cases}
\displaystyle z = - \frac{1}{\beta_0} \ln \phi
\\
\displaystyle u = a^{\frac{3}{2}} \phi
\end{cases}
\qquad \to \qquad
\begin{cases}
\displaystyle \phi = e^{-\beta_0 z}
\\
\displaystyle a = u^\frac{2}{3} e^{\frac{2 \beta_0}{3} z}
\end{cases}
\end{equation}
and, setting $\beta_0 \equiv \ell$, the new Lagrangian reads
\begin{equation}
\Lagr_{T} = -\frac{1}{8} \frac{\dot{u}^2}{u} + \frac{1}{2} \ell \dot{u}\dot{z} + \frac{1}{2} \ell^2 u \dot{z}^2 - V_0 u \;.
\label{L new scalar fieldT}
\end{equation}
Also here $z$ is the cyclic variable which permits to write the Hamiltonian as
\begin{equation}
\mathcal{H}_T = \frac{\pi_z^2}{2w} + 2\pi_z \pi_u + \frac{3}{16} u \pi_u^2 + V_0 w,
\end{equation}
where $\displaystyle \pi_z, \pi_u$ are the conjugate momenta. 

Finally, let us consider  the Gauss--Bonnet equivalent  Hamiltonian for   $\sqrt{\G}$. The symmetry generators for this case are in  Eq. \eqref{NoethersolG1}.   For $k = 2$, where we have a constant kinetic term, the Lagrangian can be written as:
\begin{equation}
\Lagr = -V_0 a^3 \phi^2 + \frac{1}{2} F_0 a^3 \sqrt{\G} \phi^2 + 2 F_0 \phi^2 \dot{a}^3 \dot{\G}) \G^{-\frac{3}{2}} - 8 F_0 \phi \dot{a}^3 \dot{\phi} \G^{- \frac{1}{2}} + \frac{1}{\ell^2} a^3 \dot{\phi}^2.
\label{lagra exampleG}
\end{equation}
The condition \eqref{inner} permits to change the minisuperspace variables from $\mathcal{S} = \{a,\phi,\G \}$ to $\mathcal{S}' = \{z,u,\G \}$ and gives rise to the same system of differential equations  as Eq. \eqref{sist1} and Eq. \eqref{sist2}, \emph{i.e}
\begin{equation}
\begin{cases}
\displaystyle \mathcal{X} z = \frac{2 \ell}{3} a \partial_a z - \ell \partial_\phi z = 1
\\
\displaystyle \mathcal{X} u = \frac{2}{3} a \partial_a u - \partial_\phi u = 0,
\end{cases}
\end{equation}
with $z$ being the cyclic variable. One possible solution  is
\begin{equation}
\begin{cases}
\displaystyle z = - \frac{1}{\ell} \ln \phi
\\
\displaystyle u = a^{\frac{3}{2}} \phi
\end{cases}
\qquad \to \qquad
\begin{cases}
\displaystyle \phi = e^{- \ell z}
\\
\displaystyle a = u^\frac{2}{3} e^{\frac{2 \ell}{3} z}.
\end{cases}
\end{equation}
Replacing the new variables $u,z$ into the Lagrangian \eqref{lagra exampleG},  we get
\begin{equation}
\Lagr_\G = \frac{\G^{-\frac{3}{2}}}{54 u} \left[27 F_0 \G^2 u^3 - 54 \G^{\frac{3}{2}} u^3 (V_0 - \ell^2 \dot{z}^2) + 
  32 F_0 \dot{\G}(\ell u \dot{z}+ \dot{u})^3 + 128 F_0 \ell \G \dot{z} (\ell u \dot{z} + \dot{u})^3\right]
\end{equation}
where, as expected,  $z$ is cyclic. By a straightforward Legendre transformation,  we find the Hamiltonian
\begin{equation}
\mathcal{H}_\G = \frac{1}{4 u^2}\left[16 \G^2 \pi_\G^2 + 8 u \G \pi_\G \pi_u  - 2 F_0  \sqrt{\G} u^4 + u^2 \pi_u^2 + 4 V_0 u^4 + 3 \cdot2^{\frac{2}{3}} u \left(\frac{\G^{\frac{3}{2}}u \pi_\G}{F_0 \ell^3} \right)^{\frac{1}{3}} \left(\pi_z + \ell \pi_u \right) \right].
\end{equation}
As final remark, we can state that if the models have the same Noether symmetries, they are dynamically equivalent.

\section{ Conclusions}
\label{conc}
We analyzed  non-minimal coupling between a scalar field and  gravity, taking into account  different  geometric invariants, namely $R$, $T$, and $\G$, the curvature, torsion, and Gauss-Bonnet scalars respectively. In all cases, the  action contains three functions of the scalar field, namely the coupling, the kinetic term and the potential. We showed that, by the  Noether Symmetry Approach,  it is possible  to fix the form of the above functions of the scalar field and solve exactly dynamics. Furthermore, it  is possible to demonstrate that if the symmetries coincide, cosmologies coming from curvature, torsion and Gauss-Bonnet gravity are equivalent. In particular, this statement holds as soon as   exponential and  power--law expansions  of the scale factor of the universe are derived as exact solutions. Interestingly, GR can be recovered  in all representations as soon as $R=-T+B$ and $f(\G)=\sqrt{\G}$.  Here $B$ is the torsion boundary term. 

As concluding remark we can say that Noether symmetries are a general paradigm by which deal with cosmologies coming from different theories of gravity. According to the present results, different theories showing the same symmetries are dynamically equivalent also if coming from different conceptual foundations. Specifically, GR and metric  theories require the Equivalence Principle, the Lorentz Invariance and so on. On the other hand, TEGR and its generalization are gauge theories invariant under the translational group.  Also if we start from these very different assumptions, if the related dynamics are governed by  the same Noether symmetries, theories are equivalent. 

In forthcoming studies, these results will be generalized to other classes of gravitational theories. 

\section*{Acknowledgements}
The Authors  are supported  by the INFN sezione di Napoli, {\it iniziative specifiche} GINGER, MOONLIGHT2 and QGSKY. 

\appendix

\begin{appendix}

\section{The Noether Symmetry Approach}
\label{noether}

The Noether Symmetry Approach is widely used to deal with cosmologies coming from different theories of gravity. For example, in
\cite{Felice, Fazlollahi:2018wmp, Sharif:2018jdj, Capozziello:2012iea, Vakili:2008ea, Capozziello:2007wc, Paliathanasis:2011jq},  the approach has been used to deal with $f(R)$ gravity. In \cite{Capozziello:2014bna, Paliathanasis:2014iva, Basilakos:2013rua, Atazadeh:2011aa, Sadjadi:2012xa, Wei:2011aa}, extended $f(T)$ TEGR models have been discussed in cosmology and spherical symmetry.  In \cite{Capozziello:2016eaz, Capozziello:2014ioa, baj, baj2},  the Noether theorem has been used  to study  $f(\G)$ and $f(R,\G)$ dynamics.
Scalar-tensor actions  have been studied in  \cite{Borowiec:2014wva, Paliathanasis:2014rja, Capozziello:2012hm, Zhang:2009mm, Capozziello:2007iu}, where  the coupling and the potential are found by symmetries.  The basic formulation of the Noether Theorem for dynamical systems and cosmology  is presented in \cite{Tsamparlis:2011zz, baj2, Dialektopoulos:2018qoe,Capozziello:1996bi}. 

For the purpose of this paper, the Noether theorem can be summarized as follows. Let us consider the  set of transformations 
\begin{equation}
\begin{cases}
\Lagr(t,q^i \dot{q}^i) \to \Lagr (\overline{t}, \overline{q}^i, \dot{\overline{q}}^i)
\\
\overline{t} = t + \epsilon \xi(t,q^i) + O(\epsilon^2)
\\
\overline{q}^i = q^i + \epsilon \eta^i(t,q^i) + O(\epsilon^2) \;,
\end{cases}
\label{transf}
\end{equation}
where $\Lagr$ is the Lagrangian of the system, $t$ an affine parameter (e.g. time) and $q^i$ the  coordinates. If such a transformation leaves the equations of motion invariant, then the condition
\begin{equation}
\left[ \xi \frac{\partial }{\partial t} + \eta^i \frac{\partial }{\partial q^i} + (\dot{\eta}^i - \dot{q}^i \dot{\xi}) \frac{\partial}{\partial \dot{q}^i} \right]\Lagr = \dot{g} - \dot{\xi} \Lagr
\label{theorem2 cosmo}
\end{equation}
holds, and the quantity
\begin{equation}
J(t,q^i,\dot{q}^i) = 	\displaystyle \xi \left(\dot{q}^i \frac{\partial \Lagr}{\partial \dot{q}^i} - \Lagr \right) - \eta^i \frac{\partial \Lagr}{\partial \dot{q}^i} + g(t,q^i) \;,
\label{conservedcosmo}
\end{equation}
is a constant of motion. 

Some remarks are necessary at this point: $i)$   the quantity in the LHS of Eq. \eqref{theorem2 cosmo} is named the \emph{First Prolongation of Noether Vector}. It is  indicated as $X^{[1]}$. It is called \emph{first} prolongation since the transformation \eqref{transf} involves the first derivative of the variables, discarding the possibility of higher-order Lagrangians. By setting $\xi = 0$, we get the non-extended Noether vector $X$, which  provides symmetries which do not depend on the coordinates transformation. In such a case, the condition \eqref{theorem2 cosmo} can be rewritten as:
\begin{equation}
\left[\eta^i \frac{\partial }{\partial q^i} + \dot{\eta}^i \frac{\partial}{\partial \dot{q}^i}\right] \Lagr = 0 \quad \to \quad J = \eta^i \frac{\partial \Lagr}{\partial \dot{q}^i} \;
\end{equation}
and it can be recast in terms of Lie derivative as 
\begin{equation}
L_X \Lagr = 0\,. 
\end{equation}
It means that the Lie derivative of a Lagrangian, containing symmetries along the flux of  vector $X$, vanishes identically. The  vector $X$ provides  the possibility to introduce  a cyclic variable into the system. To this purpose, let us consider the coordinates transformation 

\begin{equation}
q^i \to Q^i(q^j)\,,
\end{equation}
and the \emph{inner derivative} of the new variables $Q^i$, defined as:
\begin{equation} 
i_X dQ^i \equiv \delta q^j \frac{\partial Q^i}{\partial q^j}.
\end{equation}
According to these definitions, the non-extended Noether vector can be written in terms of the  variables $Q^i$ as
\begin{eqnarray}
X' = (i_X dQ^k) \frac{\partial}{\partial Q^k} + \frac{\partial  (i_X dQ^k)}{\partial t} \frac{\partial}{\partial \dot{Q}^k} . \nonumber
\label{inner}
\end{eqnarray}
Imposing 

\begin{equation}
\label{relation}
 i_X d Q^1 = 1\,,\qquad \mbox{and}  \qquad i_X d Q^i = 0\,,\qquad i \neq 1,
 \end{equation}
 the infinitesimal generator of the  variable $Q^1$ is constant and the conserved quantity is
 \begin{equation}
 J = \cdot \partial_{\dot{Q}^i} \Lagr = \pi_{Q^1}\,, 
 \end{equation}
 where  $\pi_{Q^1}$ is the conserved momentum.
 In this way, the conjugate momentum related to $Q^1$ is a constant of motion and, therefore, $Q^1$ is a  cyclic variable. Summing up, the relations \eqref{relation} allow to replace a variable with the corresponding integral of motion.   
 In summary, we search  for symmetries. If they exist,  the Lagrangian is invariant under certain transformations; we write down the generator of such transformations and, by imposing Noether's identity \eqref{theorem2 cosmo}, we get  the infinitesimal generators and the form of the unknown functions into the Lagrangian. The procedure allows to reduce dynamics and, eventually, to solve it finding out exact solutions.

\end{appendix}


\begin{thebibliography}{99}

\bibitem{Addison:2015wyg}
  G.~E.~Addison, Y.~Huang, D.~J.~Watts, C.~L.~Bennett, M.~Halpern, G.~Hinshaw and J.~L.~Weiland,
  ``Quantifying discordance in the 2015 Planck CMB spectrum,''
  Astrophys.\ J.\  {\bf 818} (2016) no.2,  132
  
\bibitem{Riess:1998cb}
  A.~G.~Riess {\it et al.} [Supernova Search Team],
  ``Observational evidence from supernovae for an accelerating universe and a cosmological constant,''
  Astron.\ J.\  {\bf 116} (1998) 1009
  
\bibitem{Bamba:2012cp}
K.~Bamba, S.~Capozziello, S.~Nojiri and S.~D.~Odintsov,
``Dark energy cosmology: the equivalent description via different theoretical models and cosmography tests,''
Astrophys. Space Sci. \textbf{342} (2012), 155

\bibitem{Copeland}
  E.~J.~Copeland, M.~Sami and S.~Tsujikawa,
  ``Dynamics of dark energy,''
  Int.\ J.\ Mod.\ Phys.\ D {\bf 15} (2006) 1753
  
\bibitem{Read:2004xc}
  J.~I.~Read and G.~Gilmore,
  ``Mass loss from dwarf spheroidal galaxies: The Origins of shallow dark matter cores and exponential surface brightness profiles,''
  Mon.\ Not.\ Roy.\ Astron.\ Soc.\  {\bf 356} (2005) 107
\bibitem{Starobinsky}
  A.~A.~Starobinsky,
  ``A New Type of Isotropic Cosmological Models Without Singularity,''
  Phys.\ Lett.\  {\bf 91B} (1980) 99.
  
  
\bibitem{Guth:1980zm}
  A.~H.~Guth,
  ``The Inflationary Universe: A Possible Solution to the Horizon and Flatness Problems,''
  Phys.\ Rev.\ D {\bf 23} (1981) 347
  
\bibitem{Linde:1983gd}
  A.~D.~Linde,
  ``Chaotic Inflation,''
  Phys.\ Lett.\  {\bf 129B} (1983) 177.
  
\bibitem{Bamba:2015uma}
K.~Bamba and S.~D.~Odintsov,
``Inflationary cosmology in modified gravity theories,''
Symmetry \textbf{7} (2015) no.1, 220-240  
  
\bibitem{Sadjadi:2013nb}
  H.~Mohseni Sadjadi,
  ``Notes on teleparallel cosmology with nonminimally coupled scalar field,''
  Phys.\ Rev.\ D {\bf 87} (2013) 064028
  
\bibitem{Gibbons:1987ps}
  G.~W.~Gibbons and K.~i.~Maeda,
  ``Black Holes and Membranes in Higher Dimensional Theories with Dilaton Fields,''
  Nucl.\ Phys.\ B {\bf 298} (1988) 741.
  
\bibitem{Capozziello:1993vs}
  S.~Capozziello, R.~de Ritis and C.~Rubano,
  ``String dilaton cosmology with exponential potential,''
  Phys.\ Lett.\ A {\bf 177} (1993) 8.
  
\bibitem{Capozziello:1993ts}
  S.~Capozziello and R.~de Ritis,
  ``Minisuperspace and Wheeler-DeWitt equation for string dilaton cosmology,''
  Int.\ J.\ Mod.\ Phys.\ D {\bf 2} (1993) 373.
  

   
  
\bibitem{Sotiriou:2008rp}
  T.~P.~Sotiriou and V.~Faraoni,
  ``f(R) Theories Of Gravity,''
  Rev.\ Mod.\ Phys.\  {\bf 82} (2010) 451
  
\bibitem{Capozziello:2011et}
  S.~Capozziello and M.~De Laurentis,
  ``Extended Theories of Gravity,''
  Phys.\ Rept.\  {\bf 509} (2011) 167
  
  
\bibitem{Amendola:1993bg}
  L.~Amendola, A.~Battaglia Mayer, S.~Capozziello, F.~Occhionero, S.~Gottlober, V.~Muller and H.~J.~Schmidt,
  ``Generalized sixth order gravity and inflation,''
  Class.\ Quant.\ Grav.\  {\bf 10} (1993) L43.
  
\bibitem{Gottlober:1989ww}
  S.~Gottlober, H.~J.~Schmidt and A.~A.~Starobinsky,
  ``Sixth Order Gravity and Conformal Transformations,''
  Class.\ Quant.\ Grav.\  {\bf 7} (1990) 893.
  
\bibitem{Allemandi}
  G.~Allemandi, M.~Capone, S.~Capozziello and M.~Francaviglia,
  ``Conformal aspects of Palatini approach in extended theories of gravity,''
  Gen.\ Rel.\ Grav.\  {\bf 38} (2006) 33.
  
  
 
\bibitem{Nojiri:2017ncd}
S.~Nojiri, S.~D.~Odintsov and V.~K.~Oikonomou,
``Modified Gravity Theories on a Nutshell: Inflation, Bounce and Late-time Evolution,''
Phys. Rept. \textbf{692} (2017), 1-104

\bibitem{Nojiri:2017qvx}
  S.~Nojiri, S.~D.~Odintsov and V.~K.~Oikonomou,
``Constant-roll Inflation in $F(R)$ Gravity,"
  Class.\ Quant.\ Grav.\  {\bf 34} (2017) 

\bibitem{Nojiri:2010wj}
S.~Nojiri and S.~D.~Odintsov,
``Unified cosmic history in modified gravity: from F(R) theory to Lorentz non-invariant models,''
Phys. Rept. \textbf{505} (2011), 59.

\bibitem{Hehl}
 F.~W.~Hehl, P.~Von Der Heyde, G.~D.~Kerlick and J.~M.~Nester,
 ``General Relativity with Spin and Torsion: Foundations and Prospects,''
 Rev.\ Mod.\ Phys.\  {\bf 48} (1976) 393.


  
\bibitem{BeltranJimenez:2019tjy}
  J.~B.~Jime\'enez, L.~Heisenberg and T.~S.~Koivisto,
  ``The Geometrical Trinity of Gravity,''
  Universe {\bf 5} (2019) no.7,  173
  
\bibitem{Arcos:2005ec}
  H.~I.~Arcos and J.~G.~Pereira,
  ``Torsion gravity: A Reappraisal,''
  Int.\ J.\ Mod.\ Phys.\ D {\bf 13} (2004) 2193
  
\bibitem{Cai:2015emx}
  Y.~F.~Cai, S.~Capozziello, M.~De Laurentis and E.~N.~Saridakis,
  ``f(T) teleparallel gravity and cosmology,''
  Rept.\ Prog.\ Phys.\  {\bf 79} (2016) no.10,  106901
  
\bibitem{Aldrovandi:2013wha}
  R.~Aldrovandi and J.~G.~Pereira,
  ``Teleparallel Gravity : An Introduction,''
  Fundam.\ Theor.\ Phys.\  {\bf 173} (2013).

\bibitem{Capozziello:2019klx}
  S.~Capozziello and F.~Bajardi,
  ``Gravitational waves in modified gravity,''
  Int.\ J.\ Mod.\ Phys.\ D {\bf 28} (2019) no.05,  1942002.
  
\bibitem{Abedi:2018lkr}
  H.~Abedi, S.~Capozziello, R.~D'Agostino and O.~Luongo,
  ``Effective gravitational coupling in modified teleparallel theories,''
  Phys.\ Rev.\ D {\bf 97} (2018) no.8,  084008
  
\bibitem{Wei:2011yr}
  H.~Wei,
  ``Dynamics of Teleparallel Dark Energy,''
  Phys.\ Lett.\ B {\bf 712} (2012) 430
  
\bibitem{Ferraro:2006jd}
  R.~Ferraro and F.~Fiorini,
  ``Modified teleparallel gravity: Inflation without inflaton,''
  Phys.\ Rev.\ D {\bf 75} (2007) 084031
  
\bibitem{Nashed:2014sea}
  G.~G.~L.~Nashed,
  ``Schwarzschild solution in extended teleparallel gravity,''
  EPL {\bf 105} (2014) no.1,  10001
  
\bibitem{Bahamonde:2016grb}
  S.~Bahamonde and S.~Capozziello,
  ``Noether Symmetry Approach in $f(T,B)$ teleparallel cosmology,''
  Eur.\ Phys.\ J.\ C {\bf 77} (2017) no.2,  107

\bibitem{Capozziello:2014mea}
S.~Capozziello, G.~Lambiase, M.~Sakellariadou and A.~Stabile,
``Constraining models of extended gravity using Gravity Probe B and LARES experiments,''
Phys. Rev. D \textbf{91} (2015) no.4, 044012  
  
\bibitem{Nojiri:2006ri}
  S.~Nojiri and S.~D.~Odintsov,
  ``Introduction to modified gravity and gravitational alternative for dark energy,''
  eConf C {\bf 0602061} (2006) 06
   [Int.\ J.\ Geom.\ Meth.\ Mod.\ Phys.\  {\bf 4} (2007) 115]
  
\bibitem{Callan:1985ia}
  C.~G.~Callan, Jr., E.~J.~Martinec, M.~J.~Perry and D.~Friedan,
  ``Strings in Background Fields,''
  Nucl.\ Phys.\ B {\bf 262} (1985) 593.
  
\bibitem{Damour:1994zq}
  T.~Damour and A.~M.~Polyakov,
  ``The String dilaton and a least coupling principle,''
  Nucl.\ Phys.\ B {\bf 423} (1994) 532
  
\bibitem{Gasperini:2002bn}
  M.~Gasperini and G.~Veneziano,
  ``The Pre - big bang scenario in string cosmology,''
  Phys.\ Rept.\  {\bf 373} (2003) 1
  
\bibitem{Capozziello:1996bi}
  S.~Capozziello, R.~De Ritis, C.~Rubano and P.~Scudellaro,
  ``Noether symmetries in cosmology,''
  Riv.\ Nuovo Cim.\  {\bf 19N4} (1996) 1.
  
\bibitem{Capozziello:1993tr}
  S.~Capozziello and R.~de Ritis,
  ``Scale factor duality and general transformations for string cosmology,''
  Int.\ J.\ Mod.\ Phys.\ D {\bf 2} (1993) 367.
  
  
\bibitem{Capozziello:2015hra} 
  S.~Capozziello, G.~Gionti, S.J. and D.~Vernieri,
  ``String duality transformations in $f(R)$ gravity from Noether symmetry approach,''
  JCAP {\bf 1601}, 015 (2016)
  
\bibitem{Capozziello:2007iu}
  S.~Capozziello, S.~Nesseris and L.~Perivolaropoulos,
  ``Reconstruction of the Scalar-Tensor Lagrangian from a LCDM Background and Noether Symmetry,''
  JCAP {\bf 0712} (2007) 009
  
  
  
 
     
  
  
\bibitem{Fine:2012ze}
  D.~Fine and S.~Sawin,
  ``Supersymmetic Quantum Mechanics and the Gauss--Bonnet-Chern Theorem,''
  arXiv:1207.2751 [math-ph].
  
\bibitem{Chern:1999jn}
  S.~S.~Chern, W.~H.~Chen and K.~S.~Lam,
  ``Lectures on differential geometry,''
  (Series on university mathematics. 1)
  
  \bibitem{GBtheorem} Y. Li "The Gauss--Bonnet-Chern Theorem on Riemannian Manifolds", 2011arXiv1111.4972L, (2011)
 
\bibitem{Easson:2020mpq}
D.~A.~Easson, T.~Manton and A.~Svesko,
``$D\to4$ Einstein-Gauss--Bonnet Gravity and Beyond,''
[arXiv:2005.12292 [hep-th]].

 
\bibitem{Zanelli:2015pxa}
J.~Zanelli,
``Chern-Simons Forms and Gravitation Theory,''
Lect. Notes Phys. \textbf{892} (2015), 289-310  

    
\bibitem{Saltas:2010ga}
  I.~D.~Saltas and M.~Hindmarsh,
  ``The dynamical equivalence of modified gravity revisited,''
  Class.\ Quant.\ Grav.\  {\bf 28} (2011) 035002
  
\bibitem{Rashidi:2018lwq}
  R.~Rashidi, F.~Ahmadi and M.~R.~Setare,
  ``Particle creation in the framework of $f(G)$ gravity,''
  Astrophys.\ Space Sci.\  {\bf 363} (2018) no.9,  196
  
\bibitem{Zhong:2018tqn}
  Y.~Zhong and D.~S\'aez - Chill\'on G\'omez,
  ``Inflation in mimetic $f(G)$ gravity,''
  Symmetry {\bf 10} (2018) no.5,  170
  
\bibitem{Paolella} 
   M.~De Laurentis, M.~Paolella and S.~Capozziello,
  ``Cosmological inflation in $F(R,\mathcal{G})$ gravity,''
  Phys.\ Rev.\ D {\bf 91} (2015) no.8,  083531
  
\bibitem{S.Silva:2018irj}
  M.~V.~d.~S.~Silva and M.~E.~Rodrigues,
  ``Regular black holes in $f(G)$ gravity,''
  Eur.\ Phys.\ J.\ C {\bf 78} (2018) no.8,  638
  
\bibitem{Myrzakulov:2010gt}
  R.~Myrzakulov, D.~Saez-Gomez and A.~Tureanu,
  ``On the $\Lambda$CDM Universe in $f(G)$ gravity,''
  Gen.\ Rel.\ Grav.\  {\bf 43} (2011) 1671
  
   
\bibitem{Li:2007jm}
  B.~Li, J.~D.~Barrow and D.~F.~Mota,
  ``The Cosmology of Modified Gauss--Bonnet Gravity,''
  Phys.\ Rev.\ D {\bf 76} (2007) 044027
  
\bibitem{Nojiri:2005jg}
  S.~Nojiri and S.~D.~Odintsov,
  ``Modified Gauss--Bonnet theory as gravitational alternative for dark energy,''
  Phys.\ Lett.\ B {\bf 631} (2005) 1
  
\bibitem{Elizalde:2010jx}
  E.~Elizalde, R.~Myrzakulov, V.~V.~Obukhov and D.~Saez-Gomez,
  ``LambdaCDM epoch reconstruction from F(R,G) and modified Gauss--Bonnet gravities,''
  Class.\ Quant.\ Grav.\  {\bf 27} (2010) 095007
  
\bibitem{baj2}
F.~Bajardi and S.~Capozziello,
``$f(\mathcal {G})$ Noether cosmology,''
Eur. Phys. J. C \textbf{80} (2020) no.8, 704


\bibitem{baj}
F.~Bajardi, K.~F.~Dialektopoulos and S.~Capozziello,
``Higher Dimensional Static and Spherically Symmetric Solutions in Extended Gauss\textendash{}Bonnet Gravity,''
Symmetry \textbf{12} (2020) no.3, 372.


 \bibitem{Manos}
  G.~Kofinas and E.~N.~Saridakis,
  ``Teleparallel equivalent of Gauss-Bonnet gravity and its modifications,''
  Phys.\ Rev.\ D {\bf 90} (2014) 084044.
  
\bibitem{Kostas}
  S.~Capozziello, M.~De Laurentis and K.~F.~Dialektopoulos,
  ``Noether symmetries in Gauss?Bonnet-teleparallel cosmology,''
  Eur.\ Phys.\ J.\ C {\bf 76} (2016) no.11,  629.
    
  
\bibitem{Capozziello:1994ie}
  S.~Capozziello, R.~De Ritis and P.~Scudellaro,
  ``Noether's symmetries in quantum cosmology,''
  Int.\ J.\ Mod.\ Phys.\ D {\bf 3} (1994) 609.
    
\bibitem{Paliathanasis:2014rja}
  A.~Paliathanasis, M.~Tsamparlis, S.~Basilakos and S.~Capozziello,
  ``Scalar-Tensor Gravity Cosmology: Noether symmetries and analytical solutions,''
  Phys.\ Rev.\ D {\bf 89} (2014) no.6,  063532
  
\bibitem{Capozziello:2013qha}
  S.~Capozziello and M.~De Laurentis,
  ``Noether symmetries in extended gravity quantum cosmology,''
  Int.\ J.\ Geom.\ Meth.\ Mod.\ Phys.\  {\bf 11} (2014) 1460004
  
\bibitem{Capozziello:1994du}
  S.~Capozziello and R.~de Ritis,
  ``Noether's symmetries and exact solutions in flat nonminimally coupled cosmological models,''
  Class.\ Quant.\ Grav.\  {\bf 11} (1994) 107.
  
    

  
\bibitem{Capozziello:1994dn}
  S.~Capozziello, R.~De Ritis and P.~Scudellaro,
  ``Noether's symmetries in nonflat cosmologies,''
  Nuovo Cim.\ B {\bf 109} (1994) 159.
  
  
  
    
  \bibitem{Felice}
  S.~Capozziello and A.~De Felice,
  ``f(R) cosmology by Noether's symmetry,''
  JCAP {\bf 0808} (2008) 016
  
  
  
\bibitem{Fazlollahi:2018wmp}
  H.~R.~Fazlollahi,
  ``F(R) cosmology via Noether symmetry and $\Lambda$-Chaplygin Gas like model,''
  Phys.\ Lett.\ B {\bf 781} (2018) 542
  
\bibitem{Sharif:2018jdj}
  M.~Sharif and I.~Nawazish,
  ``Wormhole geometry and Noether symmetry in $f(R)$ gravity,''
  Annals Phys.\  {\bf 389} (2018) 283
  
\bibitem{Capozziello:2012iea}
  S.~Capozziello, N.~Frusciante and D.~Vernieri,
  ``New Spherically Symmetric Solutions in f(R)-gravity by Noether Symmetries,''
  Gen.\ Rel.\ Grav.\  {\bf 44} (2012) 1881
  
\bibitem{Vakili:2008ea}
  B.~Vakili,
  ``Noether symmetry in f(R) cosmology,''
  Phys.\ Lett.\ B {\bf 664} (2008) 16
  
\bibitem{Capozziello:2007wc}
  S.~Capozziello, A.~Stabile and A.~Troisi,
  ``Spherically symmetric solutions in f(R)-gravity via Noether Symmetry Approach,''
  Class.\ Quant.\ Grav.\  {\bf 24} (2007) 2153
  
\bibitem{Paliathanasis:2011jq}
  A.~Paliathanasis, M.~Tsamparlis and S.~Basilakos,
  ``Constraints and analytical solutions of $f(R)$ theories of gravity using Noether symmetries,''
  Phys.\ Rev.\ D {\bf 84} (2011) 123514
  

\bibitem{Capozziello:2014bna}
  S.~Capozziello, M.~De Laurentis and R.~Myrzakulov,
  ``Noether Symmetry Approach for teleparallel-curvature cosmology,''
  Int.\ J.\ Geom.\ Meth.\ Mod.\ Phys.\  {\bf 12} (2015) no.09,  1550095
  
\bibitem{Paliathanasis:2014iva}
  A.~Paliathanasis, S.~Basilakos, E.~N.~Saridakis, S.~Capozziello, K.~Atazadeh, F.~Darabi and M.~Tsamparlis,
  ``New Schwarzschild-like solutions in f(T) gravity through Noether symmetries,''
  Phys.\ Rev.\ D {\bf 89} (2014) 104042
  
\bibitem{Basilakos:2013rua}
  S.~Basilakos, S.~Capozziello, M.~De Laurentis, A.~Paliathanasis and M.~Tsamparlis,
  ``Noether symmetries and analytical solutions in f(T)-cosmology: A complete study,''
  Phys.\ Rev.\ D {\bf 88} (2013) 103526
  
\bibitem{Atazadeh:2011aa}
  K.~Atazadeh and F.~Darabi,
  ``$f(T)$ cosmology via Noether symmetry,''
  Eur.\ Phys.\ J.\ C {\bf 72} (2012) 2016
  
\bibitem{Sadjadi:2012xa}
  H.~Mohseni Sadjadi,
  ``Generalized Noether symmetry in f(T) gravity,''
  Phys.\ Lett.\ B {\bf 718} (2012) 270
  
\bibitem{Wei:2011aa}
  H.~Wei, X.~J.~Guo and L.~F.~Wang,
  ``Noether Symmetry in $f(T)$ Theory,''
  Phys.\ Lett.\ B {\bf 707} (2012) 298
  
\bibitem{Capozziello:2016eaz}
  S.~Capozziello, M.~De Laurentis and K.~F.~Dialektopoulos,
  ``Noether symmetries in Gauss–Bonnet-teleparallel cosmology,''
  Eur.\ Phys.\ J.\ C {\bf 76} (2016) no.11,  629
  
\bibitem{Capozziello:2014ioa}
  S.~Capozziello, M.~De Laurentis and S.~D.~Odintsov,
  ``Noether Symmetry Approach in Gauss--Bonnet Cosmology,''
  Mod.\ Phys.\ Lett.\ A {\bf 29} (2014) no.30,  1450164
  

\bibitem{Borowiec:2014wva}
  A.~Borowiec, S.~Capozziello, M.~De Laurentis, F.~S.~N.~Lobo, A.~Paliathanasis, M.~Paolella and A.~Wojnar,
  ``Invariant solutions and Noether symmetries in Hybrid Gravity,''
  Phys.\ Rev.\ D {\bf 91} (2015) no.2,  023517
  
\bibitem{Capozziello:2012hm}
  S.~Capozziello, M.~De Laurentis and S.~D.~Odintsov,
  ``Hamiltonian dynamics and Noether symmetries in Extended Gravity Cosmology,''
  Eur.\ Phys.\ J.\ C {\bf 72} (2012) 2068
  
\bibitem{Zhang:2009mm}
  Y.~Zhang, Y.~g.~Gong and Z.~H.~Zhu,
  ``Noether Symmetry Approach in multiple scalar fields Scenario,''
  Phys.\ Lett.\ B {\bf 688} (2010) 13
  
\bibitem{Tsamparlis:2011zz}
  M.~Tsamparlis and A.~Paliathanasis,
  ``The geometric nature of Lie and Noether symmetries,''
  Gen.\ Rel.\ Grav.\  {\bf 43} (2011) 1861.
  
  \bibitem{Dialektopoulos:2018qoe}
  K.~F.~Dialektopoulos and S.~Capozziello,
  ``Noether Symmetries as a geometric criterion to select theories of gravity,''
  Int.\ J.\ Geom.\ Meth.\ Mod.\ Phys.\  {\bf 15} (2018) no.supp01,  1840007
 

  

\end{thebibliography}
\end{document}